\documentclass[a4paper,twoside,english]{article}
\usepackage[sumlimits]{amsmath}

\newtheorem{thm}{Theorem}[section]

\newtheorem{deff}{Definition}[section]

\def\NP{{\sf{NP}}} 
\def\BPP{{\sf{BPP}}} 
\def\BQP{{\sf{BQP}}}

\usepackage{babel}

\newcommand{\QPIP}{\textsf{QPIP}}
\newcommand{\QAS}{\textsf{QAS}}

\newcommand{\ket}[1]{|{#1}\rangle}
\newcommand{\ignore}[1]{}

\begin{document}
\author{Dorit Aharonov\footnote{School of Computer Science, The Hebrew
    University of Jerusalem, Israel. $doria@cs.huji.ac.il$. 
Supported in part by ERC-2011-StG grant 
280157-QHC, ISF grant 1446/09} and Umesh Vazirani\footnote{Department of Computer Science, UC Berkeley, California $vazirani@eecs.berkeley.edu$. Supported in part by NSF Grant CCF-0905626 and Templeton Foundation Grant 21674.}}

\title{Is Quantum Mechanics Falsifiable?  
A computational perspective on the 
foundations of Quantum Mechanics.} 

\maketitle
\begin{abstract}

Quantum computation teaches us that quantum mechanics exhibits  
exponential complexity. We argue that the standard scientific paradigm 
of ``predict and verify'' cannot be applied to testing quantum mechanics in this limit of 
high complexity. We describe how QM can be tested in this regime 
by extending the usual scientific paradigm to include {\it interactive experiments}. 

This paper is to appear in the Philosophy of Science 
anthology ``Computability: 
G$\ddot{o}$del, Turing, Church, and beyond'', Editors: Copeland, Posy, Shagrir, 
MIT press, 2012. 
\end{abstract}  

\section{Overview}
Saying that QM is paradoxical is an understatement:
Feynman once said~\cite{FeynmanL} "I think I can safely say that no one 
understands quantum mechanics". Quantum mechanics has
been a great source of fundamental issues and paradoxes in the philosophy of 
science: ranging from its statistical nature and stretching of causality
to the measurement problem. A totally new kind of 
philosophical problem arises once we 
focus on computational aspects of QM.

Indeed, the seeds of this go back to the birth of the field of quantum computation, in the demonstration that quantum computers seem to violate the 
Extended Church-Turing 
thesis. This thesis asserts that 
any reasonable computational model can be simulated 
efficiently\footnote{By efficient, we mean that the simulation overhead is 
bounded by some polynomial. i.e. $t$ steps on the computational model in question
can be simulated in $poly(t)$ steps on a probabilistic Turing machine}
by the standard model of classical computation, namely, a probabilistic 
Turing machine. Bernstein and Vazirani \cite{BV} and Simon \cite{Si}
 showed that quantum computers are capable 
of exponential speedups over classical 
models of computing, thus demonstrating the violation of the thesis 
by quantum computers~\footnote{these early 
quantum algorithms were cast in the so-called 
black-box computational model, which is a restricted model of computation in which such exponential separations can actually be proved;  
see Nielsen and Chuang \cite{nielsen}.}. 
That this violation has profound practical consequences 
became apparent shortly thereafter when Shor~\cite{shor} made his seminal 
discovery of an efficient quantum algorithm for factoring, 
one that is exponentially faster than
any known classical algorithm for this important computational problem. 
The philosophical implications of this computational view of QM are only just 
beginning to be understood, and this is the subject of this paper.

At the root of the extravagant computational power of QM is the fact that the 
state of a quantum system of $n$ spins is a unit vector in a Hilbert space of 
dimension 
that scales exponentially in $n$. This means that $2^n$ complex numbers are
required to describe the state of such a system, as compared to $O(n)$ for a 
comparable classical system. This number of parameters is larger than 
the estimated number
of particles in the universe already for $n = 500$. 
This is certainly among the most paradoxical
predictions of quantum mechanics. Is this description unnecessarily inefficient?
Could there be a different succinct description? This was a 
question posed by Feynman in his 
seminal paper~\cite{Feynman} that led to the birth of quantum computation. 
The theory of 
quantum computation provides an answer: an exponential description 
is necessary, as long as we believe that quantum computers provide an 
exponential speedup over classical computers. 
A different way to say this is that, as we believe today, 
the computational complexity exhibited by
many-body Quantum systems is 
exponentially more powerful than that of classical systems. 

One of the goals of physics research is to test the validity of 
a theory in various limits --- e.g., in the 
limit of high energy, or at the Planck scale, or close to 
the speed of light. 
Ideas from quantum 
computation point at a new regime in which to 
test quantum mechanics: the 
limit of high (computational) complexity. 
The aspects of quantum mechanics that have so far been 
experimentally verified,   
to exquisite precision (with certain predictions of QED verified to within 
one part in $10^{12}$)!) 
can all be classified as low complexity quantum mechanics - many 
of them rely on little more than single particle quantum mechanics. 
The relevant computational complexity measure 
here is the effective dimension of the Hilbert space in which the state of the system lives, and in those systems, it is very small.  
Thus a Bose-Einstein condensate, though a macroscopic quantum state, is 
effectively 
a two-level quantum system and therefore of low complexity. 
Moving beyond the practical 
difficulties in experimentally dealing with large scale quantum systems, here 
we ask whether testing QM in the high complexity limit is even theoretically 
possible, or whether there are fundamental
obstacles that prevent such testing. 

At a naive level, the issue is how can an experimentalist 
test an exponentially more powerful system than his own computational 
abilities, with the view of testing that it is indeed exponentially more 
powerful? 
Slightly more accurately, we ask: 
how can a classical, computationally restricted 
experimentalist test the high complexity aspects of quantum 
mechanics? 
The scientific method would suggest setting
up an experiment, and checking that the outcome is consistent with the prediction of the theory. 
But if the prediction of the theory requires exponential resources to compute, as we 
believe is the case for many-body quantum mechanics, is there a way to 
effectively carry out the procedure laid out in the scientific method
in order to test QM? 
On the face of it, the answer is no, since the predictions of the experiment 
cannot be computed in a reasonable time 
(of course the predictions could be efficiently computed 
using a quantum computer, but the correctness of that result would rely on 
the exponential
scaling of QM, which is what the experiment was trying to test in the first place!). Following this logic, one would deduce that the 
testing of quantum mechanics in the limit of high complexity is not 
susceptible to 
the scientific method.

A more formal way of understanding the issue is in 
terms of two ways of interpreting the 
extended 
Church-Turing thesis. 
The first interpretation
goes back to Alan Turing's conception of a Turing machine 
as an idealized model for a mathematical calculation 
(think of the infinite tape as an
infinite supply of paper, and the Turing machine control as the mathematician, or for 
our purposes a mathematical physicist calculating the outcome of an 
experiment). By this interpretation the Turing machine, i.e. the idealized 
model of a mathematical calculation, can efficiently 
simulate any other algorithmic model of 
computation. This is the interpretation which a logician might take.  
The other interpretation of the extended Church Turing thesis
reads ``reasonable model of computation'' 
as ``{\it physically realizable} model of computation''. It 
argues via the equivalence
between Turing Machines and cellular automata (which may 
be regarded as discrete analogs
of local differential equations) that Turing Machines represent, or capture, 
the evolution of physical systems
in the classical World; they can efficiently 
simulate any computational model 
that is reasonable from the point of view of physics implementation. 
Combining the two interpretations we get the following:   
in principle, any scientific theory for the classical World 
(by the second interpretation) can be formalized as a cellular automaton; 
this cellular automaton 
(by the first interpretation) can be efficiently simulated by the 
mathematical physicist to calculate the outcome
of the desired experiment. Thus, by the extended Church Turing thesis, 
the outcome of the experiment of any scientific theory can be calculated 
efficiently by the mathematical physicist. This outcome 
can then be verified experimentally, following the usual scientific paradigm. 
The source of the problem explained 
above is that in a quantum World, 
we run into a fundamental problem, since the mathematical physicist is 
still classical, whereas the quantum World 
that he is trying to understand is exponentially 
more powerful. i.e. the two interpretations of the extended Church 
Turing thesis diverge exponentially; it no longer holds that the 
mathematical physicist can calculate efficiently 
the outcome of quantum experiments. 
This is the source of the fundamental problem in testing QM in the limit of high complexity. 

Vazirani observed already in \cite{vazirani} 
that Shor's quantum factoring algorithm 
constitutes a counter example to the above 
line of thought by which the exponential computational complexity 
of quantum mechanics seems to lend it impossible to testing in the 
usual scientific paradigm. The reason is that 
Shor's algorithm can be interpreted as a suggestion for a 
layout of a physical experiment, that tests quantum mechanics in  
a regime which is believed to be impossible to efficiently simulate 
by classical computational means. Indeed, a closer examination reveals
in what way this experiment lies outside the usual scientific paradigm of 
"predict the outcome of the experiment and verify that this is the experimental 
outcome." In the case of quantum factoring, the output of the experiment consists of the prime factors of the
input $N$. The intractability of factoring on a classical computer rules out
the possibility of predicting the output of the experiment. 
Instead the verification
is performed by checking that the product of the prime factors output by the experiment is 
$N$. Thus rather than predicting the actual outcome of the experiment, what is 
predicted is that the outcome passes a
 test specified by a certain {\it computational process} 
(i.e. multiplication of the output numbers results in $N$). 
This might seem like a minor difference, between verification and comparing, 
and in a very special case; however as we shall soon see, it is
the tip of the iceberg. 

Aharonov, Ben-Or and Eban \cite{ABE} suggested that this view of 
Shor's algorithm as an experiment verifying quantum Mechanics in 
complex regimes, could be greatly generalized
by casting it in the framework of interactive proof systems. 
This is a central concept in 
computational complexity theory \cite{goldwasser1985kci, babai}
(and see \cite{arora}). 
In an interactive proof system, a computationally
weak (i.e., of polynomial strength) verifier, Arthur, 
can interact with a much more powerful (in fact unbounded) 
but untrusted entity, called 
the prover, or Merlin. By this interaction he can  
determine the correctness of an assertion made by Merlin. 
For this to be possible, 
Merlin has to be willing to answer a number of 
cleverly chosen random questions related 
to the original claim; the questions need to be random 
so that Merlin cannot prepare in advance, and thus Arthur may catch him 
if he is trying to cheat by revealing the inconsistencies in his 
claims. Arthur adaptively generates this sequence of questions based on 
Merlin's answers, and checks
 Merlin's answers for consistency. The remarkable property
of such protocols is that Arthur can only be convinced of the original claim (with non-negligible 
probability over the choice of questions) if it is in fact a valid claim. Arthur's confidence in the
claim does not depend in any way on his trust in Merlin, 
but rather in the consistency checks
that he is able to perform on Merlin's answers (see \cite{arora} for some detailed examples of such protocols). 
In complexity theory Merlin is a hypothetical
being, and the properties of 
the game between Arthur and Merlin provide deep insights 
into the nature and complexity of proofs.  
In the quantum context, \cite{ABE} suggest to replace the 
all-powerful prover Merlin by  
a real entity, namely a 
quantum system performing quantum evolutions or quantum computations 
\footnote{Note that this notion of quantum interactive proofs is 
very different from 
another notion of quantum interactive proofs
 \cite{watrous}, in which Arthur is a quantum polynomial time system 
and Merlin 
is a hypothetical all-powerful entity, 
which is studied in the literature in the 
context of quantum complexity theory.}. 

Let us understand the implications of such 
an interactive proof system in the context of a 
classical experimentalist (who is computationally ``weak'', namely limited 
to polynomial computations), who wishes to verify that the 
outcome of a quantum experiment is consistent with quantum mechanics
(which is a computationally powerful system). Let us place the 
experimentalist in the
role of Arthur, and quantum 
systems (or all of quantum mechanics) in the role of Merlin. 
Using such protocols of interaction as in interactive proofs, we will derive 
that although
the experimentalist might not be able to verify directly that the outcome of a 
{\it single}
experiment in isolation is correct 
(according to the predictions of quantum mechanics), because he would 
not be able to predict its outcome, he still could set up a sequence 
of experiments and test that the outcomes of all 
these experiments jointly satisfied the consistency checks
(mandated by the interactive proof system). 
If they did indeed satisfy them, 
he could conclude that the outcome of the original
experiment was indeed correct 
according to the predictions of quantum mechanics. Moreover, his 
confidence in this conclusion
would be based only on the success of the consistency tests, 
which he could perform efficiently. 
Of course, if the outcomes of the experiments did 
not pass the consistency tests, then 
the experimentalist could only conclude that at least one of the 
experiments failed to meet the predictions of quantum mechanics; 
the implication would be that something in the sequence of experiments 
must not fit the theory: 
either the system was not prepared correctly, 
or quantum mechanics itself is false.  

This kind of an interactive proof between the experimentalist and the 
quantum system may be 
thought of as a new kind of experiment, involving a well-designed sequence 
of interactions (a {\it protocol}) between the 
experimentalist and the system. Indeed, this would
provide a new paradigm for the scientific method, 
breaking with the "predict and test" paradigm that goes back 
several centuries. However, whether
such an interactive proof system exists for all of quantum mechanics 
is currently an open question.
Indeed, this is one of the currently most important computational
questions about the foundations of quantum mechanics.

Is it plausible that a classical 
verifier could efficiently check an exponentially more powerful 
system such as quantum mechanics? 
The earlier discussion about formulating Shor's algorithm
as an interactive experiment provides an example of the possibility to 
test QM in complex regimes; note however that the factoring problem is special,
since it is in the complexity class $NP \cap co-NP$, believed not to contain all of quantum computation, and so 
this does not clearly imply anything for testing quantum mechanics in 
general. 
Turning to more general cases, 
interactive proof systems are known to exist for systems that 
are more powerful than quantum mechanics,
for example, for the class \#P of all counting 
functions\footnote{\label{fn}to define \#P, recall for example
the satisfiability problem, which asks given
a propositional 
formula $f(x_1, \ldots, x_n)$ whether there is a satisfying assignment $a_1, \ldots, a_n$ such
that $f(a_1, \ldots, a_n) = 1$; this is a problem in \NP; 
its counting version, denoted $\#SAT$ is 
the question how many such satisfying assignments are there, out of the 
$2^n$ potential solutions.}; By these interactive proofs, the
prover can 
prove to the weak verifier that he has 
computed the answer to the \#P function 
correctly, even though such function is extremely hard for the verifier to 
compute on his own. 
However, these interactive proofs are not useful in our context, 
since the prover in them is all-powerful (or at least as powerful as \#P), 
whereas we need the prover 
to be no more powerful than QM; how would a restricted prover prove 
the correctness of a quantum mechanical evolution? 
The auxiliary random questions generated by Arthur must 
be solvable efficiently by a quantum system to 
ensure that the experiments corresponding to the interactive proof are
feasible. 
Hence, this still does not provide an answer to the above open question. 

Aharonov, Ben-Or and Eban \cite{ABE} were able to prove an interesting 
kind of interactive proof
system for quantum mechanics, which partially addresses the above question. 
In this system, Arthur, the experimentalist, 
is not purely classical, but can store and
manipulate a constant number ($3$ to be concrete) of qubits, and can exchange qubits with Merlin, who is an
arbitrary quantum system. They gave a protocol by which Arthur 
can verify that an arbitrary quantum experiment
(modeled by an arbitrary sequence of quantum gates)
has been faithfully carried out 
by Merlin, by exchanging a sequence of specially
chosen quantum messages with Merlin, the quantum system.  
At all times in this protocol, Arthur holds at most $3$ qubits. 

One way to understand this protocol is to imagine that a 
company QWave claims to have experimentally realized
a quantum computer, and wishes to convince a potential buyer that the 
computer is indeed capable of
performing an arbitrary quantum computation on up to $n$ qubits. If the 
potential buyer has the capability of 
storing and manipulating $3$ qubits, and 
of exchanging qubits with the quantum computer, 
then by following the protocol of \cite{ABE}, he 
can verify that the computer faithfully carried out any quantum computation 
of his choice. Alternatively, assume that an experimentalist
 trusts that QM describes his system of few qubits 
to extremely high precision, but does not know that that is true, 
or to what extent it is true, as the number of particles in the system 
increases. He can use the above protocol to test this,
based on his already established 
belief that his small systems obey QM to very high 
precision and a relaxed assumption about the quantum nature of the physical 
system\footnote{The assumption here is that any physical system involved, including the entire system of the prover, 
is describable by the general structure 
of quantum mechanics, namely, it can be assigned a density matrix on a tensor product space. 
One need not assume however that this larger system is coherent, or can be described 
by pure superpositions, 
or any other assumption that makes it ``fully'' quantum; for example, 
the larger systems could in principle be greatly decohered.}.
\normalsize

In the following we flesh out the main ingredients required to make 
the above line of thought rigorous; we explain how computationally weak 
(polynomial nearly classical verifier) can test the complex 
regime of quantum mechanics using interactive experiments. 

\section{Polynomial time and the Extended Church Turing Thesis}

A fundamental principle in computational complexity theory, can be summarized as equating
{\em efficient = polynomial time:} computations are considered efficient if they can be carried out
in a number of steps that is bounded by a polynomial in the size of the input. Here the size of the
input is measured in the number of bits required to specify it. This identification of efficient with 
polynomial time is to be contrasted with brute force search which takes exponential time
in the size of the input. For example, the satisfiability problem (SAT) asks whether a given propositional 
formula  $f(x_1, \ldots, x_n)$ is satisfiable. i.e. whether logical values (True and False) can be assigned 
to its variables in such a way that makes the formula true. There are $2^n$ such truth assignments, and 
brute force search over these possibilities is prohibitively expensive ---  even for $n = 500$, $2^n$ is larger than estimates for the number of particles in 
the Universe, or the age of the Universe in femto-seconds. But is this brute force search necessary?
The famous P = NP? problem asks 
whether this problem can
 be solved in a number of steps bounded by some polynomial in 
$n$.

The principle of efficient computation is also 
closely tied to the extended Church-Turing thesis, which states that any
"reasonable" model of computation can be simulated by a (probabilistic) Turing machine with 
at most polynomial simulation overhead. i.e. for any reasonable model of computation there is a polynomial $p(x)$ such that 
$T$ steps on this model can be simulated in at most $p(T)$ steps on a Turing machine. 
This means that Turing machines not only capture the notion of effective computability (which is the essence of the original
Church-Turing thesis) but they also capture the notion of efficient computation. 

As was briefly touched upon in the introduction, there are two ways to interpret what it means for
a model of computation to be "reasonable." The first may be thought of as modeling a mathematician 
carrying out a long calculation through a sequence of steps, each of which can be carried out using 
pencil and paper, and where the recipe for the sequence of steps is finitely specified. The second 
is to consider a physical model of computation or a digital computer. The computational model must be "reasonable"
in the sense that it must be physically realizable in principle. For example, implementing infinite precision
arithmetic in a single step would be considered unreasonable, since it does not account for the inevitable
noise and inaccuracy in any physical realization. Informally, one may argue that classical physics is 
described by local differential equations, which taking into account the inevitable noise and lack of infinitely precise control, 
reduces as a computational model to a cellular automaton. Since cellular automata are polynomially 
equivalent to Turing Machines, the extended Church-Turing thesis may be thought of as providing a 
constraint on what kinds of functions can be computed efficiently by Nature. 

\section{Interactive Proofs}

Let us start with a simple example. Given two graphs $G_1= (V_1, E_1)$ and $G_2= (V_2, E_2)$
we wish to test whether the two graphs are isomorphic. i.e. is there a bijection $f: V_1 \rightarrow V_2$
on the vertex sets, such that $\{u, v\} \in E_1$ iff
 $\{f(u), f(v)\} \in E_2$. i.e. edges in $G_1$ are mapped
to edges in $G_2$ under this bijection. 
There is no efficient algorithm known to solve graph isomorphism
in the worst case.

Suppose we had a powerful entity, Merlin, who claimed to be able to solve arbitrary instances of graph isomorphism. 
How could he convince Arthur about the answer to a particular instance $G_1= (V_1, E_1)$, $G_2= (V_2, E_2)$?
If the two graphs are isomorphic, then he would simply provide the bijection, and Arthur could efficiently check 
that this bijection maps edges to edges and non-edges to non-edges. If the number of vertices in each graph 
were $n$, then Arthur would need to perform $O(n^2)$ such checks.

Returning to the example, if the two graphs are non-isomorphic, how would Merlin convince Arthur
that this was the case? On the face of it, this appears impossible, since Merlin would have to rule out all
possible bijections, and there are exponentially many of those as a function of $n$. This is where the
interactive proof comes in. Arthur chooses one of the two graphs at random (according to the flip of
a fair coin), and then randomly permutes the vertices and sends the resulting description to Merlin. 
Merlin is asked to identify which of the two graphs Arthur chose. 

The point is that if the two graphs were isomorphic, then there is no way to distinguish a random 
permutation of one graph from a random permutation of the other (the two distributions are identical). 
So Merlin can succeed with probability at most $1/2$. On the other hand if the graphs were not 
isomorphic then Merlin, who can solve the graph isomorphism problem, can easily identify which
graph was sent to him, and answer accordingly. Arthur can of course easily check if Merlin answered correctly. 
Repeating this protocol $k$ times independently 
at random would decrease Merlin's probability of succeeding in convincing Arthur by chance in the case of isomorphic
graphs to at most $1/2^k$. 

A much more sophisticated protocol works in the case that Merlin claims to be able to solve the problem 
$\#SAT$, defined in footnote \ref{fn}. 
In this problem, the input is a propositional
formula $f(x_1, \ldots, x_n)$ on $n$ Boolean inputs $x_i \in \{0, 1\}$,
and the desired output is the number of 
distinct inputs to $f$ on which it evaluates to $1$. In the protocol
Arthur queries Merlin about a number of related propositional
formulae $f_1, f_2, \ldots f_m$ chosen based on $f$ and on some random 
coins; Arthur accepts only if
Merlin's answers satisfy certain consistency checks. The protocol
is fairly complicated and will not be discussed here
\footnote{it can be found in \cite{arora}; 
in fact, its discovery was one of 
the stepping stones towards one of the most 
exciting developments in theoretical computer science over the past 
two decades, namely probabilistically checkable proofs \cite{arora}}.  
The important property of this protocol however is that if Merlin lies about 
the answer to the initial problem, 
then he is forced to keep lying in order to pass the consistency tests, until
eventually he lies about a simple enough assertion 
that even Arthur can independently verify efficiently. 

\section{Interactive proofs for quantum mechanics}
The class of computational problems that can be solved efficiently (in polynomial time) on a quantum computer 
is denoted by $\BQP$. It is well known that $\BQP \subseteq \#P$, and so 
every computational problem that can be solved in polynomial time on 
a quantum computer can also be solved by a $\#P$ 
solving Merlin. At first sight the interactive proof for $\#SAT$ described 
above, which in fact works for any problem in $\#P$, would seem to 
be exactly the kind of interactive proof system we are seeking for $\BQP$. 
Unfortunately the computations 
that Merlin must perform for this protocol, namely solving 
$\#P$ problems, are (believed to be) too hard to be performed efficiently by a 
quantum computer. So even though we end up with an 
interactive proof system for $\BQP$, it is 
not one where the prover is a \BQP\ quantum machine. The major open question is whether every problem in \BQP has an
efficient interactive proof of this type, where the prover is 
a \BQP\ machine and the verifier is a polynomial time classical machine. 
If 
we denote by \BPP the class of problems solvable in polynomial time 
by a probabilistic Turing machine, this translates to the requirement 
that the verifier is a \BPP 
machine, interacting with a \BQP prover. 

\begin{deff}\label{def:QPIP} A problem $L$ is said to have a  
Quantum Prover Interactive Proof~(\QPIP\ )
if there is an interactive proof system with the following properties:
\begin{itemize}
\item The prover is computationally restricted to \BQP.
\item The verifier is a (classical) \BPP\ machine. 
\item For any $x\in L$, $P$ convinces $V$ of the
fact that $x\in L$ with probability $\ge \frac{2}{3}$ after the conversation 
between them ended (completeness). Otherwise,
when $x\notin L$ any prover (even one not following the protocol) 
fails to convince $V$ with probability higher
than $\frac{1}{3}$ (soundness). 
\end{itemize}
\end{deff}

Formally the open question mentioned in the introduction above 
can be restated as asking whether $\BQP \subseteq \QPIP$, i.e., whether 
any problem in \BQP has a quantum prover interactive proof system 
as above; or alternatively, whether the correctness of the outcome of 
a polynomial time quantum 
evolution can be proven to a classical \BPP verifier by a \BQP prover. 
Note that the other direction trivially holds: 
$\QPIP \subseteq \BQP$, since any $\QPIP$ protocol can be simulated 
by a $\BQP$ machine
by simulating both prover and verifier, as well as the interaction 
between them; hence the question can be written as whether 
$\BQP=\QPIP$. 

Aharonov, Ben-Or and Eban \cite{ABE} managed to show a somewhat weaker result. 
To this end, they defined a relaxation of $\QPIP$: 
the verifier in their definition is  
a hybrid quantum-classical machine. Its classical part is
a $\BPP$ machine as above. 
The quantum part is a register of $c$ qubits (for some
      constant $c$ - $3$ would suffice), on which the verifier
 can perform arbitrary quantum operations, as well as send them to the 
prover, who can in its turn perform further operations on those qubits 
and send them back. At
      any given time, the verifier is not allowed to possess   
      more than $c$ qubits. The
      interaction between the quantum and classical parts is the usual one: the
      classical part controls which operations are to be performed on the
      quantum register, and outcomes of measurements of the quantum register can
      be used as input to the classical machine.
There are two communication channels between the prover and the verifier: 
the quantum one in which the constant 
number of qubits can be sent, and the 
      classical one in which polynomially many bits can be sent. 

Aharonov, Ben-Or and Eban \cite{ABE}
 proved that with this relaxation of $\QPIP$,
which we will denote here by $\QPIP^*$,  
a \BQP\ prover can convince the verifier of any language he can
compute: 

\begin{thm} \label{thm:main}
$\BQP\ \subseteq \QPIP^*$.
\end{thm}
where the other direction, $\QPIP^* \subseteq \BQP$, is again trivial. 

The meaning of this result is that if the verifier trusts 
that his hybrid classical system with the aid of a constant number of qubits 
acts according to his quantum mechanical description of it 
with sufficient confidence, he can also be convinced with very high 
confidence of the results of the computation of the most general and 
complicated quantum computation which takes polynomial time, run on a 
quantum computer he has no control over. 
 
The mathematical method that was used to prove this result is taken 
from the realm of cryptography and uses computer science notions such as 
error correction and authentication; we will attempt to sketch it in the 
next section. 

\section{How weak verifiers can test strong machines: Proof idea}

As a warm-up to the $\QPIP^* = \BQP$ question, 
consider the following simple scenario:
suppose the verifier wishes to store an $n$ qubit state
$|\phi\rangle$ that she is about to receive, 
but she has only a constant number of qubits of quantum memory. 
Fortunately she can 
use the services of a quantum memory storage company. The problem 
is that she does not trust the company.  
Does the verifier have a way of checking that the storage company eventually
returns to her the same state that she sent them? We can think of this problem as the problem of 
trivial-computation by an untrusted prover: the verifier just wants to 
verify that the company, here viewed as the prover, 
did nothing to the state she wants to maintain, 
or, in other words, that the company applied to it the {\it identity} 
computation. The prover wishes to convince the verifier that indeed the state 
he returns is exactly the state sent by the verifier (regardless of whether this is indeed 
the case).  
The challenge is that the verifier cannot 
measure the state before storing it at the prover's hands, 
since this will collapse the state; hence the protocol needs to work even if 
she knows nothing about the 
state she stores! in fact, when she receives or generates the state, 
she cannot even hold all its qubits in her laboratory at once. 

The idea for solving this problem is in fact quite simple: 
Let us assume that
the verifier generates the unknown
quantum state $|\phi\rangle$ one qubit at a time, 
and transmits each qubit to the quantum storage company, or the prover, 
as she generates it. 
Imagine just for the sake of explanation that $|\phi\rangle$
is a tensor product state 
$|\alpha\rangle\otimes |\beta\rangle\otimes...$. 
Before sending the first qubit in the state $|\alpha\rangle$
to the company, the verifier 
encodes it into the state of, say, two qubits, in the following 
way. The verifier adds one auxiliary qubit in the 
known state $|0\rangle$, which will serve as a {\it check} qubit. 
The state at the verifier's hands after that is
$\ket{\alpha}\otimes \ket{0}$. She then chooses a 
random unitary $U_R$ uniformly from all 
the unitaries acting on two qubits, 
and applies it to her two qubits. 
She derives the state $U_R(\ket{\alpha}\otimes\ket{0})$; note that 
this is a completely random two-qubit state.  
She sends this state to the company, and then receives the next qubit
in the state $\ket{\beta}$, encodes it into another random state 
by adding an additional auxiliary qubit and applying a random (newly chosen) 
unitary on $\ket{\beta}\otimes \ket{0}$, sends the resulting 
state, and so on. 
By the end of this protocol, the company have received two
times as many qubits as the original 
state. 

To retrieve the quantum state from the company, 
the verifier asks for pairs of qubits, block by block. On each such 
block of two qubits, she first applies  
the inverse of the random unitary, $U_R^{-1}$ she applied on that 
block at the time of encoding. She then applies a measurement of  
{\it only the check qubit}, and tests that 
it is indeed in the state $\ket{0}$; she need not measure 
the first qubit in each block, and this way she does not disturb its state.  

Why does this test work with high probability? 
The point is that the initial state of each block, before applying 
the random unitary, lives inside a two-dimensional manifold $M$, 
described by all states of the form 
$\ket{\gamma}\otimes\ket{0}$ (for all one-qubit states 
$\ket{\gamma}$). 
The random unitary $U_R$ takes the manifold $M$ and maps it 
onto a random two dimensional manifold $M_R$ inside the four dimensional 
space of the two qubits. Since the company has no information regarding 
$U_R$, from its point of view, the resulting manifold $M_R$ is a completely 
random two dimensional manifold inside this four dimensional space. 
Therefore, if the company tries to alter the state without being caught, 
or if the company 
cheated, and in fact, it does not have a reliable quantum memory, the final 
state of the two qubits will be, with 
extremely high probability, significantly far away from that manifold, 
since the company knows nothing about this manifold. 
When the verifier performs the test of rotating the 
state of the two qubits back by the inverse unitary (thus rotating the 
two dimensional manifold $M_R$ back to the original one $M$)
the state of the auxiliary qubit, which should be 
$\ket{0}$, will be 
significantly far away from it;   
when the verifier checks that the 
additional qubit is in the state $\ket{0}$, she
 will detect that such a change had occurred with high probability. 

Notice that in the above explanation we never used our assumption 
that the initial state is a tensor product state; in fact, 
everything holds even in the presence of multi-particle entanglement 
between all different qubits of the original state. 
This idea thus already contains the seed for the 
solution we are looking for, because it enables checking the ability to 
{\it store} highly complicated states, namely to perform the identity 
computation on them. This is true even though the verifier herself 
does not have large enough quantum memory capabilities to even hold the state, 
and moreover, she doesn't know which state she is trying to store.  

This notion of maintaining complicated quantum states by untrusted 
parties, and the ability 
to detect whether they were altered,   
was in fact invented in the context of cryptography, and is called 
a {\it quantum authentication scheme} (\QAS) \cite{barnum2002aqm}.
It turns out that the quantum authentication scheme described above is 
not so useful, since to realize it, we need to
efficiently select a random unitary, which involves infinite accuracy 
issues. Moreover, 
the security of the above scheme is yet to be proven; 
technically, this is non-trivial due to the
continuous nature of this scheme. 
\cite{ABE} describe a slightly more involved 
 \QAS\ whose security is much easier to prove since it is amenable to 
a terminology which describes errors in the state as discrete. 

To move to the language of discrete errors, 
recall the notion of Pauli matrices: these are the 
four $2X2$ matrices 
\begin{equation}\label{pauli}
I=\left( \begin{array}{cc}
1 & 0 \\
0 & 1 \end{array} \right), 
\sigma_x=\left( \begin{array}{cc}
0 & 1 \\
1 & 0 \end{array} \right),
\sigma_y=\left( \begin{array}{cc}
0 & i \\
-i & 0 \end{array} \right),
\sigma_z=\left( \begin{array}{cc}
1 & 0 \\
0 & -1 \end{array} \right)\end{equation}
acting on one qubit. 
These four matrices linearly 
span the continuum of changes that can happen to one qubit.  
It thus suffices to handle the probability that one of the non-identity 
operators $\sigma_x,\sigma_y, \sigma_z$ occurs to a qubit to be able to 
provide bounds on the probability 
that the state had been altered significantly while 
at the hands of the company. 

We now modify the above described \QAS\ as follows. 
The encoding is done essentially as before, by adding one qubit in the state 
$\ket{0}$ to each qubit in the original state, and applying 
a random rotation on those two qubits. However,  
instead of choosing the rotation randomly from all possible unitary operations
on two qubits, as in the first \QAS\ , the random unitary $U_R$ is  
chosen from a finite subgroup of all two-qubit unitaries, 
called the {\it Clifford group}. This group is defined to be the 
group of unitaries $C$ such that when $C$ acts 
on a Pauli matrix $P$ by conjugation, 
(namely, $P$ is mapped to $CPC^{-1}$), the resulting matrix is 
still inside the Pauli group. In other words, the Pauli group is closed to 
conjugation by matrices from the Clifford group. 
The reason this property is advantageous is 
this: imagine that when at the hands of the company, the
 matrix $E$ is applied to a quantum state $\psi$, rather than the identity. 
The effect on the {\it density matrix} $\rho$ describing the state
is that $E$ acts on it by conjugation, so $\rho$ is mapped to $E\rho E^{-1}$. 
Now, if a random Clifford $C$ 
is applied to the state before the state is transformed to the 
hands of the company, and then its reverse is applied 
when the state is returned, the effective overall 
action on $\rho$ is $C^{-1}(E(C\rho C^{-1})E^{-1})C=
(C^{-1}EC)\rho (C^{-1}E^{-1}C)$. Thus, the effect of first rotating $\rho$ 
by a Clifford operator (and then rotating back),
is that not $E$ was applied, but its
conjugation by a random Clifford matrix. Recall that 
we can span $E$ in terms of 
Paulies; it turns out that applying a random 
Clifford before $E$ is applied, makes $E$ effectively equal to 
the application of a random Pauli matrix to each qubit independently, 
and this includes the extra test qubit.   
At verification stage such a procedure
will go undetected by the verifier's measurements only if the Pauli
on the check qubit turns out to be either identity or $\sigma_z$, 
but if it is $\sigma_x$ or $\sigma_y$ it will be detected since the 
measured state will be $\ket{1}$. 
This happens with probability $0.5$.  

Hence, it turns out that 
even though the restriction to the Clifford group seems
quite strong, (in particular, the Clifford group contains a finite number of 
elements rather than a continuum!) this choice provides enough 
randomization to completely 
randomize the action of the prover, so that any tampering of the 
state will be detected with high probability. 

We therefore have a solution to the simple problem of maintaining 
a quantum state using an untrusted storage device. 
It might seem that this problem is far too simple, and almost irrelevant 
for the more general problem of verifying that a quantum circuit 
had been applied correctly on the state.  
In fact, the above idea suffices to solve the entire problem by a simple 
modification. 
Imagine now that we have a certain computation, or a specification of a 
quantum algorithm by a sequence of two-qubit quantum gates, $U_1,....,U_T$. 
We want to make sure that with high probability, the final state the prover 
sends us is very close to the correct state, $U_T\cdots U_1|0^n\rangle$. 
Given the \QAS\ above
we can achieve this goal as follows.
The verifier starts by encoding each qubit in the input state using 
the Clifford based \QAS\ above. 
She then sends these pairs of qubits to the 
prover one pair at a time. All that the prover needs to do is  
serve as an untrusted storage device! 
The verifier is the one to perform the gates one by one, and this is 
done as follows: 
to apply the $i$th gate, the verifier asks the prover to send her the 
two blocks of two qubits encoding the relevant qubits on which she 
wants to apply the gate (assuming all gates act on two qubits).
Upon receiving those $4$ qubits, she first decodes them, by applying the 
inverse of the random unitaries to each block 
(of course she must maintain a classical description of those unitaries in 
her classical memory so she can carry out these steps). 
After decoding she applies the gate $U_i$ on the two decoded qubits, and encodes 
the two qubits again using two freshly chosen random unitaries;  
she then sends the two blocks to the prover. This way she had updated 
correctly the 
state at the hands of the prover; if the prover does not tamper with it, 
after time step $i$ he holds an authenticated version of the correct state 
of the quantum circuit after $i$ gates.  
After all gates have been applied this way, the prover (if honest) holds the authenticated 
version of the final state of the quantum circuit. 
At this point 
the verifier asks the prover to send her all pairs of qubits 
one by one; once she receives such a pair, she applies the decoding using the 
inverse of the relevant encoding unitary, and then measures the test qubit in the block 
to check that it is in the 
state $\ket{0}$; if all tests pass successfully, 
she is assured (with constant probability) 
that the state she had received is correct. 
Of course, her constant probability can be amplified by repeating the entire process. 

The size of the quantum register which the verifier is assumed to 
possess in this protocol is
$4$ qubits, since the length of the encoding of one qubit is $2$. 
In fact, $3$ qubits suffice, since  
it suffices to send one register at a time and
wait until the verifier decodes it before sending the
second register of two qubits. 

\section{Further complications: Can the prover apply the gates?} 
The Clifford based \QAS\ enables us to prove the main theorem, but there are
several problems with this proof. 
First, it is not fault tolerant, namely, 
if the storage devices and the quantum channels between the prover 
and verifier are faulty and 
not ideally isolated from the surrounding, then the scheme will 
quickly fail. Fault tolerance is necessary for the scheme to be applicable
in a physically realistic setting. But a 
more severe criticism can be made against 
the above scheme: it seems like cheating, as 
the computation is performed entirely by the verifier! 
In terms of making progress towards the $\QPIP = \BQP?$ question, 
it is crucial that the prover perform the non-trivial quantum computation, 
since eventually we want the verifier to be classical. What we want is 
that the prover apply the quantum gates, {\em without} 
learning any information about the encoded 
qubits that he is applying the gates to; 
because if he has such information, he can use it to alter the state 
of the qubits from the correct state, without being caught. 
Hence, we need some way of encoding the qubits that are delivered to 
the prover, so that a) he does not know the encoding  b) 
he is capable of performing gates on the encoded state nevertheless. 
Though the above two requirements might seem contradictory, 
it is possible to achieve both at the same time. 

At this point it is not known how to achieve both goals using 
the Clifford based \QAS\, since it does not seem to have sufficient 
structure.
Aharonov et al. \cite{ABE} achieve this by providing a second \QPIP\ protocol,
based on a 
different \QAS\ ,due to to Ben-Or, Cr\'epeau, Gottesman, Hassidim and Smith
\cite{benor2006smq}. We explain this in this section. 
The result is thus a \QPIP\ in which the verifier does not perform any 
quantum computation except for the encoding in the beginning; moreover, 
the interaction with the prover, after those encoded qubits are sent, 
is completely classical.  

The idea of how such a manipulation can be carried over is inspired 
by the notion of error detection codes from computer science. 
An error detection code 
is a mapping of a string of, say, $k$ bits 
$s\in \{0,1\}^k$ into a longer string of $m$ bits, $w\in \{0,1\}^m$. 
We say this code detects $t$ errors if whenever $t$ bits or fewer have flipped 
in $w$ this fact can be detected. 
This can be achieved of course only if there 
is sufficient redundancy in the encoding, i.e., $m$ is sufficiently larger than  
$k$. 

Quantum error detection codes are defined similarly:
the encoding is a unitary embedding 
of the space of quantum states of $k$ qubits into the space of quantum 
states of $m$ qubits. The notion of 
bit-flips as errors in the classical case 
is replaced by non-identity Paulies applied to
the qubits. More precisely, in the quantum setting we are faced with 
quantum errors of the following types: a bit flip, described by the Pauli 
$\sigma_x$ (see Equation \ref{pauli}), a phase flip, described by the Pauli matrix 
$\sigma_z$, and a combination of both, described by $\sigma_y$.   

One can generalize the notion of quantum error detecting codes 
also to qudits, namely to states of higher dimensional particles, say, each of dimension 
$q$; instead of Pauli matrices, one considers 
their generalizations: $\sigma_x$ is generalized by the operator 
$X_q:\ket{x}\mapsto \ket{x+1 ~ mod ~ q}$, which we refer to as a generalized 
bit flip; $\sigma_z$ is generalized 
by $Z_q:\ket{x}\mapsto e^{2\pi i x/q}\ket{x}$, which we refer to as a generalized phase flip, 
and 
the generalized Pauli group over $F_q$, applied to one qudit, 
is the group generated by $X_q$ and $Z_q$, namely, 
it consists of the set of combinations of those errors, 
$X_q^\ell Z_q^n$ for $\ell, n\in \{0,\cdots,q-1\}$. 

Why do we discuss error detection codes in our context? 
The requirement for error detection resembles the requirements from 
a \QAS\, in which we need to be 
able to detect {\it any} modification induced by the prover to the string 
of qubits we would like him to maintain. The difference is 
that the task we are aiming at in the context of \QAS\ is more difficult: 
in error detection codes the goal is to detect any error which tampers at most 
$t$ (qu)bits, and if more than $t$ (qu)bits have been tampered, 
we do not care if this 
goes undetected. By contrast, in \QAS\
we make no assumption on the number of locations which can be tampered with; 
we have no control over how the prover can 
alter the state, and we would like to be able to detect 
{\it any} error. 

It is not difficult to convince one's self that 
no error detection code can achieve this ambitious goal of being able 
to detect any error. However, 
if we relax our requirements and 
allow the detection of errors only  
with high probability rather than with certainty, this becomes possible!
The main idea behind the authentication scheme of \cite{benor2006smq}
is to use not a fixed error detection code but one which is chosen randomly
from a large set of possible codes. 
In this way, \emph{any error} 
(namely, any combination of non-identity 
Paulies) is detected by all but a small
fraction of codes that can be used, and so, with high probability, 
any error will be detected. 

This randomized encoding resembles
the usage of a random unitary in the previous two \QAS\.  
The difference is that in the scheme we will now describe, 
due to the algebraic structure of the codes 
which we will use, the prover will be able to apply gates on the states
without knowing which of the error detection codes is being used.  
We now explain how this is done. 

We start with a familiar quantum error detection code, called 
the polynomial code \cite{FT}. 
We operate over the field $F_q$, for a large prime $q$. 
Recall that a polynomial $f=\sum_{i=0}^d a_i x^i$ of 
degree $d$ over this field is 
determined by its $d+1$ coefficients $a_i$. Imagine now we represent this 
polynomial in terms not of its coefficients, but rather its values 
at different points of the field, say, $1 \cdots m$. 
In other words, we encode the polynomial by the string of values 
$f(j)$ for $j=1 \cdots m$. 
If $m$ is larger than $d+1$, then there is redundancy in this 
representation and errors can be detected. In fact, 
if we pick $m=2d+1$, then exactly all errors which contain up to 
$d$ altered locations can be detected. 
The above scheme gives a classical code (and 
a very famous one - it is the Reed-Solomon code \cite{arora}). 

To get a quantum detection code, we need to be 
able to detect not only classical errors (the generalization of 
$\sigma_x$ to quantum systems of dimension $q$, namely $X_q$ and its powers 
$X_q^\ell$) 
but also quantum errors, a.k.a as generalized phase
flips (the generalizations of $\sigma_Z$ to quantum systems of dimension $q$, 
$Z_q$, and its powers $Z_q^n$.). 
To achieve this, we consider {\it superpositions} of all polynomials
of degree up to $d$.  
We encode each element $a$ in the field $F_q$ by the following state: 

\[\ket{S_a}=\sum_{f, deg(f)\le d, f(0)=a}\ket{f(1),...,f(m)}.\]

\normalsize
Namely, by the superposition of all strings evaluating a polynomial of degree less than $d$ 
at $m$ points in the field, where the sum is only over such polynomials whose value at $0$ 
is $a$. 
That such a superposition possesses ability to detect classical-like errors
of weight up to $d$, namely errors of type $X_q$ and their powers
applied to at most $d$ coordinates, 
follows from the fact that each string in the 
superposition possesses it, due to the classical properties of
error detection of polynomial codes. The ability to detect $d$ phase flips 
is less straight-forward. It 
follows from changing the basis and looking at the state after 
the application of the Fourier transform. On one hand, $Z_q$ is transformed in this basis to 
$X_q$; so being able to correct for generalized bit flips in the new basis translates to 
correcting phase flips in the original basis.
The point is that after applying the Fourier transform, 
we arrive the dual of the superposition of all polynomials, which 
turns out to be 
also a superposition of those polynomials! Hence, in this basis too we can detect
bit flips, which means correcting phase flips in the original basis. 
Since bit flips and phase flips, and their combinations, span the entire unitary group, 
this means that we are able to detect any error if it did not involve too many locations.   
This gives us the polynomial quantum code, which can thus 
be shown to detect errors at up to $d$ locations, 
namely applications of any combination of 
the non-identity generalized Paulies
on at most $d$ locations. 

So far, we have defined only one error detection code, which we refer to  
as our {\it basic} code. We would now like to introduce a randomization process, 
which would result 
in a {\it family} of error detection codes, such that any error 
applied by the prover will be detected with high probability. 

If we assume that the prover only applies a Pauli operator at each 
location, then it suffices to randomize the basic code as follows: 
For each location, 
a random sign flip $\epsilon_i$ (plus or minus) is chosen 
independently. $\epsilon=(\epsilon_1,\cdots,\epsilon_m)$ is called 
the {\it sign key}. 
The basic code is modified according to $\epsilon$ by 
multiplying each location by $\epsilon_i$, and the result is called the 
{\it signed polynomial code}: 

\[\ket{S_a|_\epsilon}=\sum_{f, deg(f)\le d, f(0)=a}
\ket{\epsilon_1f(1),...,\epsilon_mf(m)}.\]

It turns out that this randomization suffices to ensure that the verifier 
detects with high probability any (non-trivial) 
Pauli group operator applied by the prover on {\it any number} of locations.
This is because in order for it not to be 
detected, the Pauli applied on top of the sign key
has to match the values of a low degree polynomial, and this 
happens with small probability. 

The above scheme does not suffice, however, to handle 
a general operator applied by the prover. 
To this end \cite{benor2006smq} use an additional random key,
which they call the {\it Pauli 
key}; for each location, the verifier not only picks a random sign, 
plus or minus,  
but also a random element of the generalized Pauli group $X_q^{x_i}Z_q^{z_i}$.  
The Pauli key is denoted $(x,z)=((x_1,\cdots,x_m),(z_1,\cdots,z_m))$.   
To encode, the verifier applies the generalized Pauli 
$X_q^{x_i}Z_q^{z_i}$ 
on the $i$th coordinate of the state $\ket{S_a|_\epsilon}$ 
encoded by the random signed 
polynomial code, for all $i$; this derives the state 
$$\ket{S_a|_{\epsilon,x,z}}=(X_q^{x_1}Z_q^{z_1})\otimes (X_q^{x_2}Z_q^{z_2})\otimes\cdots \otimes 
(X_q^{x_m}Z_q^{z_m}) \ket{S_a|_\epsilon}.$$ 
It is technically not too difficult to see that 
due to symmetry arguments, this randomization by a random Pauli key makes 
the general operator that the prover had applied
appear effectively as though the prover applied a uniformly chosen 
random Pauli operator at each 
location; from there the argument proceeds as before, to show that this is 
indeed a \QAS\ which can detect any error with high probability. 

It remains to explain how the prover can apply 
gates on the encoded state even though
he does not know the encoding, namely, he has no information
regarding the Pauli and the sign keys, $(x,z)$ and $\epsilon$.   
The idea is that the prover perform the gates 
assuming that the basic polynomial code was used, 
namely that $\epsilon$ and $(x,z)$ where trivial. 
The verifier and prover then need to perform very simply corrections 
on top of that. Essentially, the verifier needs to
update his Pauli and sign keys for each gate the prover applies,    
where for some gates, the prover needs to measure part of his state 
and send the classical result to the verifier, in order for the 
verifier to know how to update his keys.
We demonstrate the details with two examples of gates, 
but this maneuver can be done for a universal set of quantum gates. 

The first example is very simple. 
Imagine that 
the prover is required to apply the gate $X_q:\ket{a}\mapsto \ket{a+1 mod q}$, 
on the encoded state, namely, it is required to take 
$\ket{S_a|_{\epsilon,x,z}}\mapsto \ket{S_{a+1}|_{\epsilon,x,z}}$. 
This can be achieved if the prover applies 
the operations $X_q^{\epsilon_i}$ to the $i'th$ coordinate, for all $i$; 
To see this, first check the case in which both $\epsilon$ and 
$(x,z)$ are trivial; Indeed, in this case, the state is 
simply $\ket{S_a}$ and it is easy to check that applying 
$X_q$ on each coordinate, namely, adding one to every coordinate of the 
polynomial translates  
to adding $1$ to the value of the polynomial at $0$, namely $a$; this
is the exact operation we wanted. 
The more general claim, that for general $\epsilon$ applying 
$X_q^{\epsilon_i}$ achieves the desired 
result, is not much more difficult. 
 
However, the prover cannot apply 
$X_q^{\epsilon_i}$ on the $i$th coordinate
since he does not know $\epsilon$... So how can the prover apply the gate? 
Fortunately, the prover need not do anything to apply the gate! 
Instead, all that needs to be done is that the {\it verifier} 
updates his Pauli key by decreasing $\epsilon_i$ from $x_i$, 
 $x_i \mapsto x_i-\epsilon_i$; this 
effectively achieves the same result. 

The second example is just slightly more complicated: 
it is the application of the Fourier transform gate, 
$F\ket{a}\mapsto \frac{1}{\sqrt{q}}\sum_{b=0}^{q-1} e^{2\pi i ab/q}\ket{b}$. 
Once again, when the keys are trivial, 
it is easy to check that if the prover applies 
the gate $F$ on each coordinate in $\ket{S_a}$, 
the total effect is the desired Fourier transform gate 
$\ket{S_a}\mapsto \frac{1}{\sqrt{q}}\sum_{b=0}^{q-1} e^{2\pi i ab/q}\ket{S_b}$.  
But we need to correct for the existence of the Pauli and sign keys. 
First, we observe that the sign key does not change anything in the above 
argument - that code too is mapped to itself (i.e., it is self-dual) 
by the coordinate-wise application of the Fourier transform \cite{benor2006smq}.  
As for the correction because of the Pauli key, 
observe that the conjugation by the Fourier transform maps 
$X_q$ to $Z_q$ and $Z_q$ to $X_q^{-1}$, 
and so $Z_qF=FX_q$  and $X_q^{-1}F=FZ_q$.  
Hence, the prover can apply the Fourier gate on each coordinate, 
and the verifier need only correct his Pauli key 
from $(x,z)$ to $(-z,x)$. 
 
In a very similar way, the prover can perform a universal set of
gates without knowing the encoding, namely the sign key or the Pauli key.  
We proceed just as in the previous scheme: the verifier applies
the gates with the help of the prover, as above, one by one, 
and at the end, the verifier checks for correctness by asking the prover 
to send him the blocks of encoded qubits, and checking that each block 
lives in the code space of the correct random error correcting code. 

The intriguing fact about this scheme is that the prover can be manipulated to apply the gates 
without knowing the encoding. This raises hope that similar methods might 
be applicable even when the verifier is entirely classical, 
and thus, that a completely classical verifier can be convinced of the 
correctness of a QM evolution. 

Finally, we remark regarding the reasons this scheme enables fault tolerance.  
First, the fact that the prover can apply gates on his own, makes 
it possible for him to apply gates in parallel on all the qubits he 
maintains in his memory;   
In the usual noise
model, in which qubits are faulty even when no gates are applied to them (this
model is known as ``faulty wires''), error
correction must be applied constantly on a constant fraction of the qubits
\cite{aharonov1996}. Since the quantum space of the verifier is limited in
our model, the prover must be able to perform those error corrections,
and thus must perform many gates in parallel. 
The proof that this scheme is fault tolerant relies on standard 
quantum fault tolerance
proofs, for example, \cite{FT}, but some additional 
care is required, since the verifier can only hold a constant number 
of qubits at a time, while he is the only one who can authenticate 
qubits. 


\section{Summary}
The standard scientific paradigm going back several hundred years --- 
"predict and verify" --- is not sufficiently powerful to test quantum mechanics
in the high complexity regime. The exciting possibilities 
and challenges that this regime poses call for an extension 
of the scientific paradigm to interactive experiments. Such interactive 
experiments, inspired by the notion of interactive proofs from computer 
science, allow testing 
complex and powerful systems without the need to be able to predict their 
behavior; the example described here provides a way to do so using a 
classical verifier with very small scale quantum capabilities. 
Whether or not a completely classical verifier can test 
quantum evolutions is left open. 

A number of issues call for further thought. A very difficult question is whether
there is a reasonable straw man theory that agrees with current experimental
data about QM but {\em does not} violate the extended Church Turing thesis. 
A theory of confirmation based on interactive experiments remains to be 
developed; important initial steps in this direction were taken in \cite{JonathanThesis}.
Another direction to explore is whether such a theory of interactive experiments
might be useful for testing other highly complex systems besides those suggested by QM.
It is intriguing to further understand the philosophical foundations of 
a theory of confirmation in which interaction between a weak verifier and 
a highly complex physical system can take place.

\section{Related Work}
Initiated independently, and followed by discussions with us, 
Jonathan Yaari has studied  
the notion of ``interactive proofs with Nature'' in his thesis in 
Philosophy of science~\cite{JonathanThesis}. 
His thesis provides an initiation of the study of a theory of confirmation 
based on interactive proofs. 

In the context of blind quantum computation, 
Broadbent, Fitzsimons, and Kashefi's \cite{broadbent2008ubq}
suggest a protocol that provides a possible way of showing that BQP = QPIP* 
where the verifier needs only a single quantum bit. At this point it is unclear 
whether the security of this protocol can be rigorously established. 

\section{Acknowledgements}

U.V. would like to thank Les Valiant and D.A. would like to thank Oded Goldreich, Madhu Sudan,  
Guy Rothblum, Gil Kalai, and the late Itamar Pitowsky for stimulating discussions
related to the ideas presented here.



\end{document}